\documentclass[
 reprint,
superscriptaddress,
showpacs,
 amsmath,amssymb,
 aps,
prb,
floatfix,
]{revtex4-1}

\usepackage{graphicx}
\usepackage{dcolumn}
\usepackage{color}
\usepackage{bm}
\usepackage{hyperref}
\graphicspath{{Images/}}

\begin{document}

\preprint{APS/123-QED}

\title{Magnetic and electronic structure of the layered rare-earth pnictide EuCd$_2$Sb$_2$ }
\author{J.-R. Soh}
\affiliation{Department of Physics, University of Oxford, Clarendon Laboratory, Parks Road, Oxford OX1 3PU, UK}%
\author{C. Donnerer}%
\affiliation{London Centre for Nanotechnology, University College London, London WC1H 0AH, UK}%
\author{K. M. Hughes}%
\altaffiliation[Present address: ]{The Blackett Laboratory, Department of Physics, Imperial College London, London SW7 2AZ, UK}%
\affiliation{Department of Physics, University of Oxford, Clarendon Laboratory, Parks Road, Oxford OX1 3PU, UK}
\author{E. Schierle}%
\affiliation{Helmholtz-Zentrum Berlin fur Materialien und Energie, Albert-Einstein-Stra\ss e 15, D-12489 Berlin, Germany}%
\author{E. Weschke}%
\affiliation{Helmholtz-Zentrum Berlin fur Materialien und Energie, Albert-Einstein-Stra\ss e 15, D-12489 Berlin, Germany}%
\author{D. Prabhakaran}
\affiliation{Department of Physics, University of Oxford, Clarendon Laboratory, Parks Road, Oxford OX1 3PU, UK}%
\author{A. T. Boothroyd}
\email{andrew.boothroyd@physics.ox.ac.uk}
\affiliation{Department of Physics, University of Oxford, Clarendon Laboratory, Parks Road, Oxford OX1 3PU, UK}%
\date{\today}

\begin{abstract}
Resonant elastic X-ray scattering (REXS) at the Eu $M_5$ edge reveals an antiferromagnetic structure in layered EuCd$_2$Sb$_2$ at temperatures below $T_\textrm{N}$ = 7.4\,K with a magnetic propagation vector of $(0,0,1/2)$ and spins in the basal plane. 
Magneto-transport and REXS measurements with an in-plane magnetic field show that features in the magnetoresistance are correlated with changes in the magnetic structure induced by the field. 
\textit{Ab initio} electronic structure calculations predict that the observed spin structure gives rise to a gapped Dirac point close to the Fermi level with a gap of $\Delta E \sim 0.01$\,eV. The results of this study indicate that the Eu spins are coupled to conduction electron states near the Dirac point. 


\end{abstract}

\pacs{75.25.-j, 78.70.Ck, 71.15.Mb, 71.20.-b} %
\maketitle


\section{introduction \label{sec:level1}}

Condensed matter systems which combine non-trivial electron band topology and magnetic order
provide an arena {\textcolor{black}{in which} to investigate the interplay between the physics of strong electron correlations
and large spin-orbit coupling (SOC)~\cite{Armitage2016,Rau2016,Hasan2010,Pesin2009,Witczak-Krempa2014}. The intrinsic symmetries in these crystal structures afford
protection to exotic quasi-particle excitations which possess a whole host of desirable properties
such as extremely high mobility and large magnetoresistance~\cite{Liu2014,Wang2013}. 

Systems in which topological charge carriers are coupled to magnetism have strong potential for spintronic device applications, where the current can be modulated by altering the spin structure with an externally applied field. The coexistence of the two phenomena can be realized in europium based antimonides, EuX$_2$Sb$_2$ (X=Cu, Pd, Zn, Cd)~\cite{Anand2015,Das2010,weber2006low,May2012, Schellenberg2011}. These 122 pnictides possess 4$f$ electrons which give rise to strong electron correlations, and heavy Sb which produces large SOC. 

Within the family of EuX$_2$Sb$_2$, EuCd$_2$Sb$_2$ has recently garnered interest due to the discovery of a large thermoelectric figure of merit $ZT$ of 0.60 at 617\,K\cite{Zhang2010}.  This led to systematic investigations of various substitutions in CaCd$_2$Sb$_2$, Yb$_{x}$Eu$_{1-x}$Cd$_2$Sb$_2$ (Ref.~\onlinecite{Zhang2010a}), Eu(Zn$_{1-x}$Cd$_{x}$)$_2$Sb$_2$ (Ref.~\onlinecite{B916346H}) and  Eu(Cd$_{1-x}$Mn$_{x}$)$_2$Sb$_2$ (Ref.~\onlinecite{Min2015}). The large $ZT$ in EuCd$_2$Sb$_2$ was attributed to the heavy masses of Cd and Sb, which give rise to the low thermal conductivity, and to the presence of conducting Eu 4$f$ states, which leads to a large enhancement in the density of states near $E_\textrm{F}$ compared to CaCd$_2$Sb$_2$\cite{Zhang2010}. The latter feature, however, is contradicted by an electron-spin resonance (ESR) study of EuCd$_2$Sb$_2$, which suggested a localized moment picture for the Eu spins\cite{Goryunov2012}. Highly localized 4$f$ orbitals usually host strong electron--electron correlations {\textcolor{black}{which} were not included in the electronic structure calculations in Ref.~\onlinecite{Zhang2010}. Furthermore, the physics of large SOC in the Cd and Sb bands was not explored.


Consistent with strong electronic correlations, EuCd$_2$Sb$_2$ displays antiferromagnetic (AFM) order of the Eu spins below $T_\textrm{N} \simeq$ 7.4\,K, as evidenced by magnetic susceptibility~\cite{Schellenberg2011,Goryunov2012,Artmann1996,Zhang2010}, M\"{o}ssbauer spectroscopy~\cite{Schellenberg2011}, ESR~\cite{Goryunov2012} and heat capacity measurements~\cite{Zhang2010}. Moreover, electronic transport measurements show an anomaly in the conductivity at $T_\textrm{N}$, suggesting that the spin structure is coupled to the charge carriers~\cite{Zhang2010}. Up to now, however, there are no reports on the ground state magnetic structure in EuCd$_2$Sb$_2$, which would shed light on the nature and consequences of this coupling. 

In the light of this, we set out in this study (i) to determine the magnetic structure by single-crystal soft X-ray resonant magnetic scattering, and (ii) to investigate the nature of electrical conduction through \textit{ab initio} electronic structure calculations including correlations. We chose X-rays rather than neutrons for the diffraction study because of the very strong neutron absorption of both Eu and Cd.  We find that in zero field the Eu spins order in an A-type AFM structure with the spins lying in the $ab$ plane, and predict that this AFM structure gives rise to a gapped Dirac point along the $\Gamma-A$ high symmetry line in the Brillouin zone. We also find that field-induced changes in the magnetic order are correlated with features in the magneto-resistance.

\section{Experimental and theoretical methods}
\begin{figure}[t]
	\includegraphics[width=0.5\textwidth]{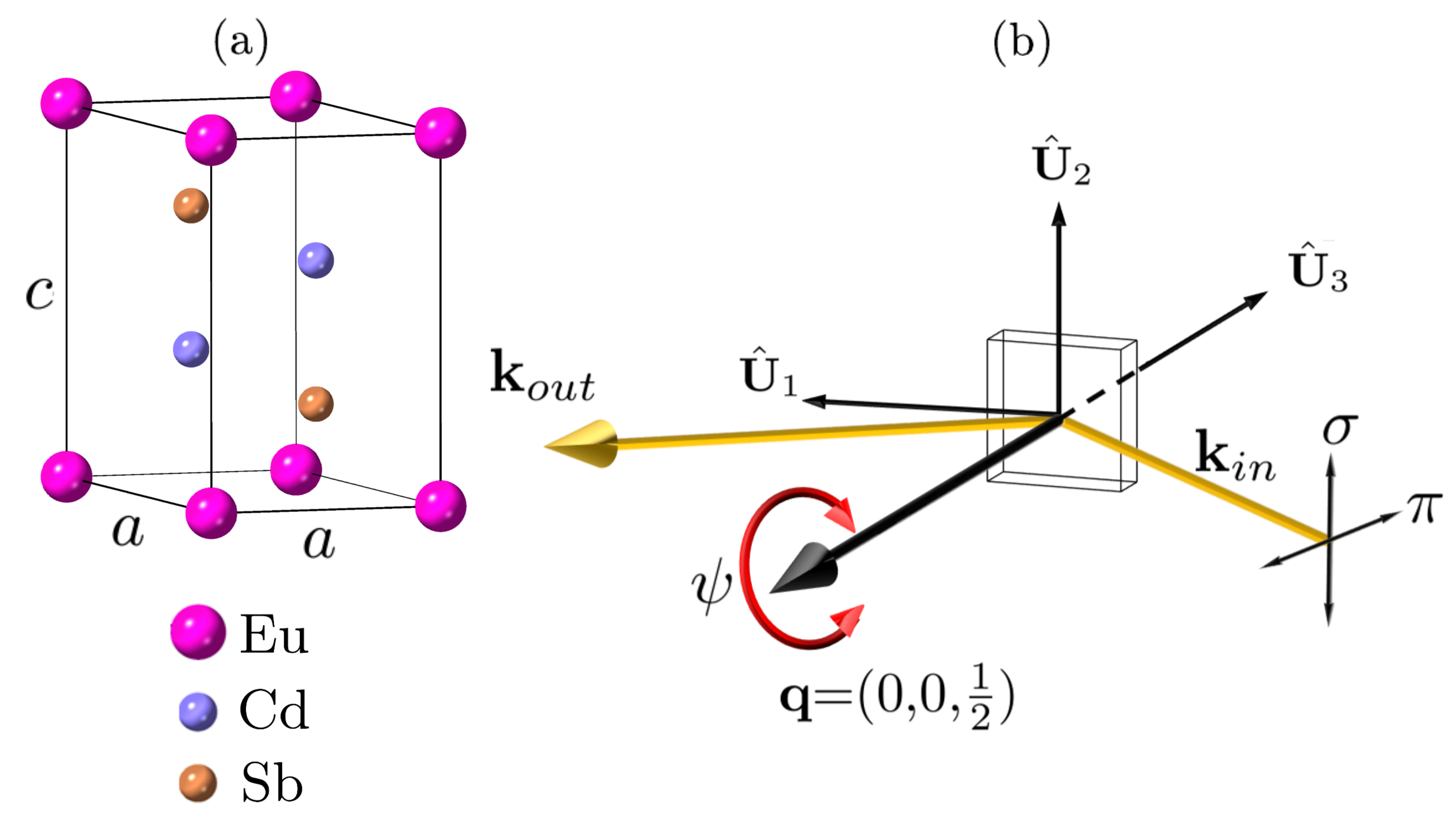}
	\caption{\label{fig:AG11b_Experimental_Set_Up} (a) A unit cell of EuCd$_2$Sb$_2$. (b) The experimental set-up for the REXS experiment, with $\hat{\textbf{U}}_1$, $\hat{\textbf{U}}_2$ and $\hat{\textbf{U}}_3$ defined as in Refs.~\onlinecite{Hill1996, Paolasini2007}. The crystal was mounted with the $c$-axis parallel to the scattering vector \textbf{q}. In the high magnetic field chamber, the field was applied along the direction of $\hat{\textbf{U}}_1$ ($H \perp c$). The magneto-transport and magnetization measurements reported in this study were also performed with $H \perp c$. }
\end{figure}
EuCd$_2$Sb$_2$ single crystals were prepared by a chemical vapour transport method. All handling was carried out in an argon glove box. Stoichiometric amounts of high purity Eu (99.9\%), Cd (99.99\%) and Sb (99.999\%) elements were mixed and loaded in an alumina crucible, which was then sealed in an evacuated quartz tube.  The tube was slowly heated to 1275\,K, kept for 24\,h and quenched to room temperature.  The quartz tube was opened, and the powder reground and reloaded in the crucible with iodine as the transport agent. The crucible was sealed in the quartz tube, which was heated to 1325\,K and kept for a week before being cooled slowly to room temperature.  Small single crystals were separated from the crucible and used for our measurements. The crystal structure and crystallographic quality of the crystals was studied on a Rigaku SuperNova single-crystal diffractometer operated with a Cu $K_\alpha$ source. 

The REXS measurements {\textcolor{black}{were performed} on the UE46-PGM1 beamline at the BESSY II storage ring~\cite{ENGLISCH2001541}. A plane grating monochromator was used to tune the X-ray energy to match that of the Eu $M_5$ edge (1.1284 \,keV). The dipole transition at the $M_5$ edge (3$d$ to 4$f$) directly probes the magnetic $4f$ states, which leads to a strong enhancement of the magnetic scattering.

The REXS experiment was performed in the horizontal scattering geometry in the two-circle XUV diffractometer [Fig.~\ref{fig:AG11b_Experimental_Set_Up}(b)]. The sample was cooled below $T_\textrm{N}$ by a liquid helium flow cryostat in conjunction with an aluminium shield to reduce beam heating. The cryostat achieved a base temperature of 4.5\,K, but due to beam heating we estimate the sample temperature to be about 5\,K. As in a typical magnetic X-ray scattering experiment, the magnetic structure is determined by studying the azimuthal dependence of the scattered intensity with $\sigma$ and $\pi$ incident photons~\cite{Detlefs2004,Blume1988,Mazzoli2007,Detlefs2012}. The intensity of the scattered beam was estimated with a AXUV100 avalanche photodiode with no polarization analysis.Therefore, when $\sigma$-polarization was used, both the $\sigma \rightarrow \sigma'$ and $\sigma \rightarrow \pi'$ channels contributed to the scattered intensity. Likewise, $\pi \rightarrow \sigma'$ and $\pi \rightarrow \pi'$ channels contributed to the scattered intensity when $\pi$-polarization was used. To study the evolution of the magnetic structure in an external magnetic field, diffraction measurements were performed in the high field chamber with fields of up to 2 T applied in the direction of $\hat{\textbf{U}}_1$ [Fig.~\ref{fig:AG11b_Experimental_Set_Up}(b)].

A superconducting quantum interference device (SQUID) magnetometer (Quantum Design) was used to perform magnetization measurements as a function of temperature and field applied perpendicular to the $c$ axis ($H \perp c$). Data for $H\parallel c$ have been reported previously\cite{Goryunov2012}. Measurements as a function of field were performed for $H\leq 3$~T  at fixed temperatures $T$ between 1.8 and 8\,K, while temperature-dependent measurements were performed for $T = 1.8$ to 50\,K at fixed fields up to 1\,T. Field- and temperature-dependent electronic transport measurements were performed for $H\leq 5$ T and 2 K $\leq T \leq$ 100 K on a Physical Property Measurement System (Quantum Design). Gold wires were bonded to the single crystal with silver paste in a four probe configuration. 

To further explore the effects of magnetic order on electronic transport in EuCd$_2$Sb$_2$ we carried out \textit{ab initio} electronic structure calculations using the implementation of density functional theory (DFT) provided by the plane wave basis Quantum Espresso suite~\cite{Giannozzi2009}. A Monkhorst--Pack mesh~\cite{Monkhorst1976} of $8\times8\times6$ was used for \textbf{k}-point sampling. Relativistic pseudo-potentials were used to account for the strong spin-orbit interaction in cadmium and antimony~\cite{Hemstreet1993}. The generalised gradient approximation (GGA) functional was used to describe the exchange correlation~\cite{Perdew1996}. A spin-polarized calculation was also implemented to account for the possible spin splitting in the electronic bands due to the magnetic europium ions~\cite{kubler2017}. To model the strong electron correlations in the highly localized europium $4f$ orbitals, a correction to the exchange-correlation functional was implemented~\cite{Anisimov1991,Anisimov1993,Anisimov1997,Liechtenstein1995,Kotliar2004}. This additional functional has an associated parameter $U$, which resembles the $U$ in the Hubbard model. In this work $U$ = 3.1 eV was used, the justification for which will be given in the results section. The unit cell was doubled along the $c$-axis to accommodate the $(0, 0, 1/2)$ AFM propagation vector found in the REXS study. 

\section{Results and analysis}
Laboratory single crystal X-ray diffraction revealed the high crystalline quality of the flux-grown crystals (see Supplementary Materials [\onlinecite{Soh2018}]), and confirmed the $P\bar{3}m1$ space group with room temperature cell parameters $a=4.7030(9)$\,\AA, $c=7.7267(18)$\,\AA, and Wyckoff positions 1a $(0,0,0)$, 2d $(1/3,2/3,0.6322(3))$ and 2d $(1/3,2/3,0.2473(5))$ for the Eu, Cd and Sb atoms, respectively, in close agreement with the earlier studies\cite{Schellenberg2011,Zhang2010,Zhang2010a,Artmann1996} (see Table~\ref{tab:table1}).   These structural parameters were used in the subsequent DFT calculations.

\begin{table}[t]
	\caption{\label{tab:table1}%
		Magnetic and structural parameters of EuCd$_2$Sb$_2$ found in this work and previous investigations.
	}
	\begin{ruledtabular}
		\begin{tabular}{lllllc}
			\textrm{$T_\textrm{N}$ [K]}&
			\ \ \textrm{$\theta_\textrm{p}$ [K]}&
			\textrm{ $\mu_\textrm{eff}$}&
			\ \ \textrm{$a$ [\AA]}&		
			\ \ \textrm{$c$ [\AA]}&	
			\textrm{Reference}\\[2pt]
			\colrule
			7.4(1)  &  $-3.8(2)$  & 8.07(5) & 4.7030(9)  & 7.7267(18) & this work\footnotemark[1] \\
			7.4  &  $-4.6(5)$  & 8.11(1) & 4.699(2)  & 7.725(2) & \onlinecite{Schellenberg2011}\footnotemark[1]\footnotemark[2] \\
			7.22  &  $-3.14(7)$  & 7.83(4) & 4.6991(1)  & 7.7256(2) & \onlinecite{Zhang2010}, \onlinecite{Zhang2010a}\footnotemark[2] \\
			7.8  &  $-3.3$  & 7.37 & 4.698(1)  & 7.723(1) &  \onlinecite{Artmann1996}\footnotemark[1] \\
			7.4  &  $-3$  & 7.83 & -  & - &  \onlinecite{Goryunov2012} \footnotemark[2] \\ 
		\end{tabular}
	\end{ruledtabular}
	\footnotetext[1]{Single crystal}
	\footnotetext[2]{Polycrystal}
\end{table}

\subsection{Magnetization and magneto-transport}
\begin{figure}[b]
	\includegraphics[width=0.5\textwidth]{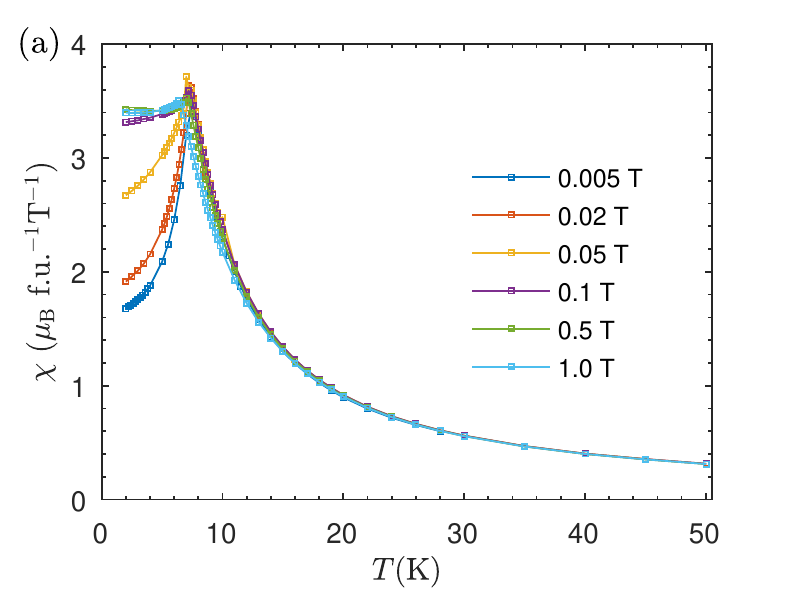}
	\includegraphics[width=0.5\textwidth]{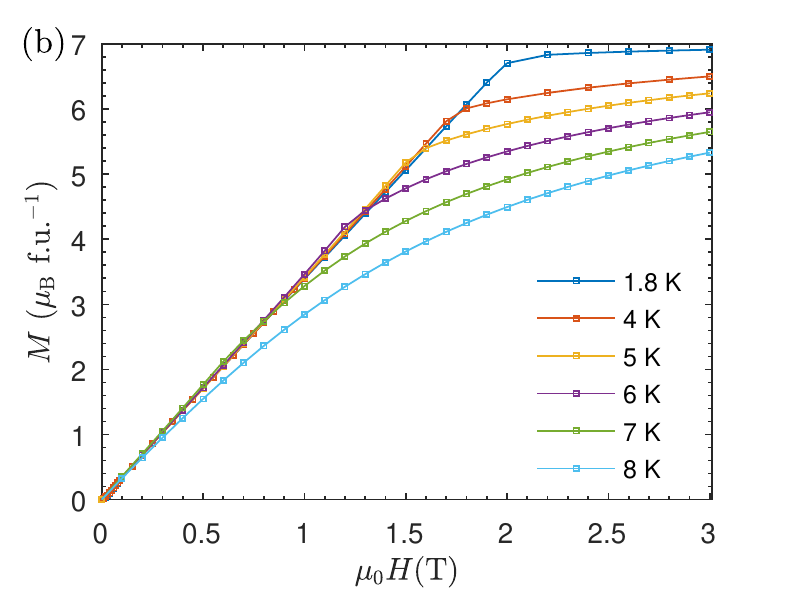}
	\caption{\label{fig:AG1e_EuCd2Sb2} (a)  Temperature dependence of the susceptibility $\chi$ measured at various field strengths with $H\perp c$. (b)  Isothermal magnetization with $H\perp c$ at various temperatures. }
\end{figure}
The temperature dependent susceptibility, $\chi(T)$, for various fields is shown in Fig.~\ref{fig:AG1e_EuCd2Sb2}(a). Upon cooling, the susceptibility first increases then peaks sharply at $T_\textrm{N} \simeq $ 7.4\,K, signalling that AFM order has set in. The low field data are well described by the Curie--Weiss law,  $\chi(T)= C/(T-\theta_p)$, $C = \mu_0\mu_\textrm{eff}^2/3k_\textrm{B}$, with $\mu_\textrm{eff} = 8.07(5)$\,$\mu_\textrm{B}$ and $\theta_\textrm{p} = -3.8(2)$\,K (see Supplementary Materials [\onlinecite{Soh2018}]). 
	
Figure~\ref{fig:AG1e_EuCd2Sb2}(b) displays the isothermal magnetization curves at various temperatures.  At $T=1.8$ K ($T \ll T_\textrm{N}$) the magnetization increases linearly with field before saturating at $M_\textrm{sat}\simeq 6.95\,\mu_\textrm{B}$\,f.u.$^{-1}$ for fields above $\mu_0H_\textrm{c} \simeq 2$\,T. As there is one Eu ion per formula unit, the values of $M_\textrm{sat}$ and $\mu_\textrm{eff}^2 = g_J^2J(J+1)\mu_\textrm{B}^2$ are fully consistent with divalent Eu$^{2+}$ ($4f^7$, $S = 7/2$, $g_J = 2$). The deviations from linearity that demarcate the AFM phase vanish above the N\'{e}el temperature. Taken together with the susceptibility, transport and REXS measurements (see below), we propose the $\mu_0H$--$T$ phase diagram given in Fig.~\ref{fig:BTdep}. The values of $T_\textrm{N}$, $\theta_\textrm{p}$ and $\mu_\textrm{eff}$ found here are consistent with those from earlier investigations, see Table~\ref{tab:table1}.
\begin{figure}[t!]
\includegraphics[width=0.5\textwidth]{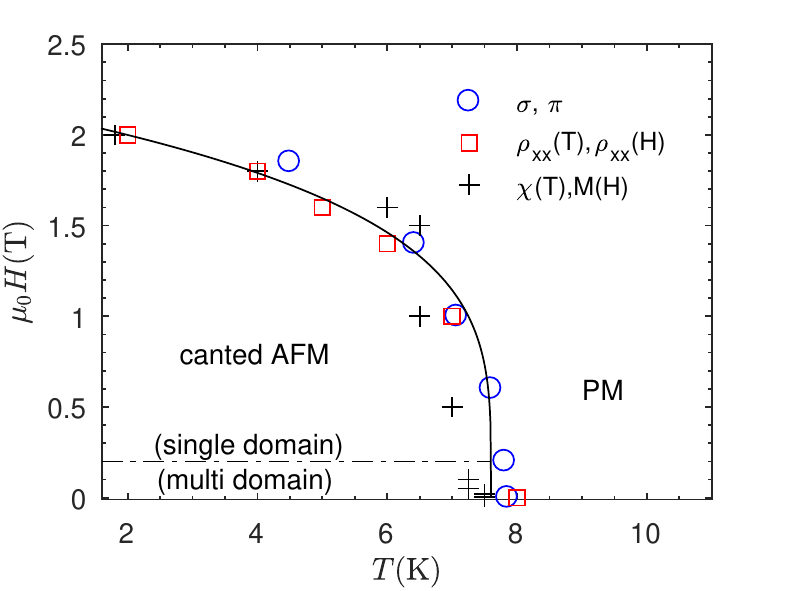}
	\caption{\label{fig:BTdep} The $\mu_0H$-$T$ phase diagram ($H \perp c$) obtained from the anomalies in the soft X-ray, magneto-transport and magnetic susceptibility data. The line demarcates the paramagentic (PM)  from the canted AFM phase. The spin scattering in the charge carriers is supressed when EuCd$_2$Sb$_2$ develops spontaneous magnetic order or when the spin structure is in the fully polarized state. A small field of 0.2 $T$ is sufficient to re-orientate the multi-domain AFM to a single-domain AFM phase with spins canted towards the applied field.}
\end{figure}

\subsection{Resonant X-ray magnetic scattering}

When the sample was cooled below $T_\textrm{N}$, a reflection with scattering vector $\textbf{q} = (0,0,1/2)$ was observed. The intensity of the peak was strongly enhanced when the photon energy was tuned to the Eu $M_5$ edge. These observations are consistent with magnetic Bragg scattering from an AFM structure in which the Eu spins are ferromagnetically aligned in the $ab$ plane and antiferromagnetically stacked along the $c$ axis, i.e.~an A-type AFM.  
\begin{figure}[t]
	\includegraphics[width=0.5\textwidth]{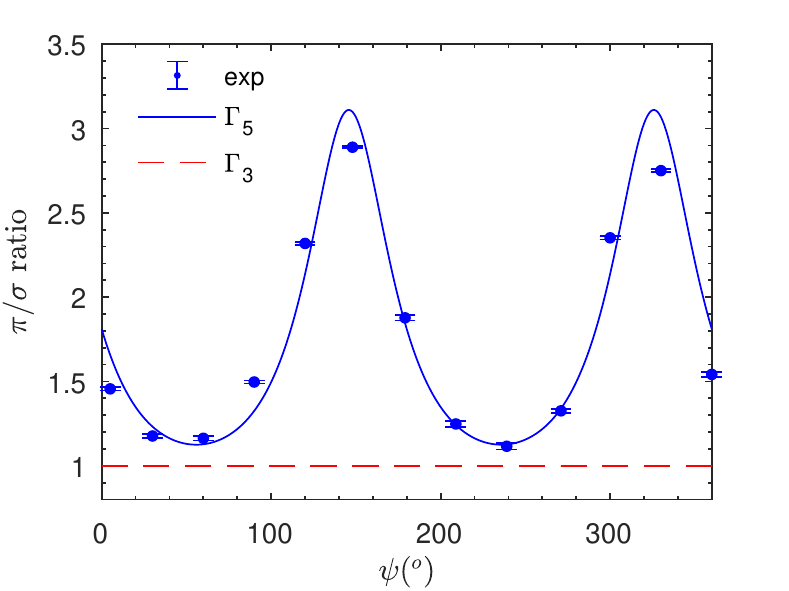}
	\caption{\label{fig:Azi_HVrat} Variation in the intensity ratio for $\pi$ and $\sigma$ incident polarizations in an azimuthal scan at the $(0,0,1/2)$ magnetic Bragg peak. The data were recorded at $T \simeq $ 5\,K. The full line shows a fit to the three domain model with Eu spins lying in the plane ($\Gamma_5$ structure). If the Eu spins point along the $c$-axis ($\Gamma_3$ structure), the $\pi$/$\sigma$ intensity ratio should be unity and show no angular dependence (broken line).}
\end{figure}
There are two irreducible representations (irreps) of the space group that are compatible with the observed propagation vector of $(0,0,1/2)$: $\Gamma_3$ and $\Gamma_5$. The magnetic structures described by these irreps differ only in the direction of the Eu spins, which point either parallel ($\Gamma_3$) or perpendicular  ($\Gamma_5$) to the $c$ axis, as shown below in Figs.~\ref{fig:AG11_Dirac_point}(a) and~\ref{fig:AG11_Dirac_point}(d). To establish which of the two irreps describes the symmetry in the AFM phase of EuCd$_2$Sb$_2$ we performed an azimuthal ($\psi$) scan and recorded the scattered intensity with both $\sigma$ and $\pi$ incident polarizations. During the scan the sample temperature was $T\simeq 5$\,K.

The calculated angular variation of the X-ray scattering amplitude for the $\Gamma_3$ and $\Gamma_5$ structures is given in Table~\ref{table:EuCd2Sb2_scattering_channels}. Results for the four linear incident and scattered polarization channels are listed, but as the polarization of the scattered photons was not analysed in our experiment the intensities in the $\sigma'$ and $\pi'$ channels for a given incident polarization need to be summed. If the spin structure has $\Gamma_3$ symmetry, the scattered intensity averaged over both final polarization states is the same for both polarization states of the incident photons. Furthermore, in our scattering geometry the $c$ axis, and hence the spins, lies along the scattering vector \textbf{q}, such that a rotation of the sample about \textbf{q} in an azimuthal scan will produce no variation in the scattering amplitude.

\begin{figure}[t]
	\includegraphics[width=0.5\textwidth]{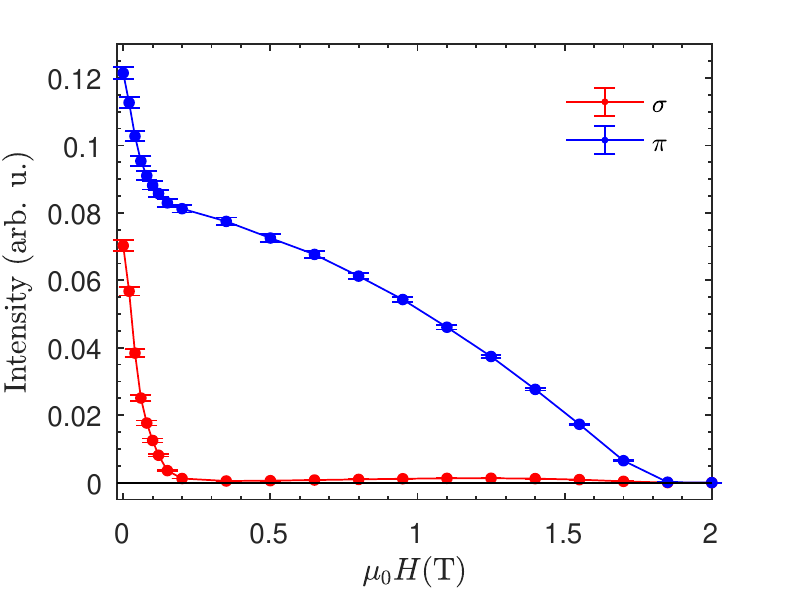}
	\caption{\label{fig:Bdep3} The field dependence ($H \perp c$) of the $(0,0,1/2)$ magnetic reflection measured with $\sigma$ and $\pi$ incident polarizations. The data were recorded at temperature $T\simeq 5$\,K and azimuthal angle $\psi = 0$.}
\end{figure}

On the other hand, if the spin structure displays $\Gamma_5$ symmetry, the scattered intensity with $\pi$ incident polarization should be larger than that with $\sigma$ polarization. This arises from the additional contribution from the  $\pi \rightarrow \pi'$ scattering channel [see Table~\ref{table:EuCd2Sb2_scattering_channels}]. Furthermore, because the Eu spins lie in the plane in this irrep, a rotation of the sample about the scattering vector \textbf{q} will produce a $\psi$ dependence in the intensity of the scattered beam.

\begin{table}[b]
	\caption{\label{table:EuCd2Sb2_scattering_channels}%
		 Calculated scattering amplitudes for the $\Gamma_3$ or $\Gamma_5$  magnetic structures of EuCd$_2$Sb$_2$ (Refs.~\onlinecite{Hill1996,lovesey1996}). $z_1$, $z_2$ and $z_3$ are the projections of the magnetic moment on to the $\hat{\textbf{U}}_1$, $\hat{\textbf{U}}_2$ and $\hat{\textbf{U}}_3$ basis vectors as defined in Fig.~\ref{fig:AG11b_Experimental_Set_Up} and Ref.~\onlinecite{Hill1996}. 
     Here $2\theta$ is the scattering angle. }
	\begin{ruledtabular}
		\begin{tabular}{ccc}
			\textrm{Scattering Channel}&\multicolumn{2}{c}{Scattering Amplitude}\\
			   & $\Gamma_3$ &  $\Gamma_5$    \\
			\colrule
			$\sigma \rightarrow \sigma'$ & 0 & 0                    \\ 
			$\sigma \rightarrow \pi'$    & $z_3 \sin \theta$ & $z_1 \cos \theta $    \\
			$\pi    \rightarrow \sigma'$ & 0 &  $-z_1 \cos \theta $ \\
			 $\pi    \rightarrow \pi'$    &   $z_3 \sin \theta$ & $-z_2 \sin 2\theta $\\
		\end{tabular}
	\end{ruledtabular}
\end{table}

In Fig.~\ref{fig:Azi_HVrat} we plot the ratio of the scattered intensity with $\pi$ incident polarization to that with $\sigma$ incident polarization, as a function of $\psi$. The data show that for all $\psi$ angles the $\pi/\sigma$ ratio is greater than 1, consistent with the $\Gamma_5$ irrep. Moreover, there is a strong  angular dependence of the $\pi/\sigma$ ratio which can be described by a three-domain model for the  $\Gamma_5$ structure, with the in-plane orientation of Eu spins in each domain rotated by $\pm~120^\circ$ around the $c$ axis relative to the other two domains. The three-domain model was recently found to describe the magnetic order in EuCd$_2$As$_2$, which has the same $\Gamma_5$ magnetic structure as found here~\cite{Rahn2018}. The domain populations that give the best fit to the data are 18.4\%, 73.5\% and 8.1\%, respectively. This preferential population of one domain could potentially arise from the fact that the angular dependent measurements were performed after the field dependent measurements below $T_\textrm{N}$.

When a magnetic field is applied in the $\hat{\textbf{U}}_1$ direction [see Fig.~\ref{fig:AG11b_Experimental_Set_Up}(b)], the scattered intensity in the $(0,0,1/2)$ reflection initially decreases strongly with field for both incident polarizations, as shown in Fig.~\ref{fig:Bdep3}. At $\mu_0H \simeq 0.2$\,T the intensity measured with $\sigma$ polarization has dropped to zero, and intensity with $\pi$ polarization has decreased by about one-third relative to zero field. At higher fields, the $\pi$ intensity continues to decrease, eventually vanishing when $\mu_0H \simeq 1.8$\,T, a field close to the critical field $H_\textrm{c}$ at which the magnetization saturates [see Fig.~\ref{fig:AG1e_EuCd2Sb2}(b)].

The field-dependent behaviour can be understood as follows. Application of small in-plane fields causes the spins in each 120$^\circ$ domain to rotate away from the field while remaining in the plane and antiferromagnetically coupled along the $c$-axis. By the time $\mu_0H \simeq 0.2$\,T, the spin component $z_1$ along the field direction  has become zero, so the intensity with $\sigma$ incident polarization vanishes and the intensity with $\pi$ incident polarization is reduced because only the  $\pi \rightarrow \pi'$ channel contributes (Table~\ref{table:EuCd2Sb2_scattering_channels}). Fields above $0.2$\,T induce a FM component along the field direction which saturates at $\mu_0H \simeq 1.8$\,T. This canted spin structure can be regarded as a combination of FM order of the $z_1$ spin components and AFM order of the $z_2$ spin components. As a result, FM Bragg peaks appear and increase in intensity with increasing field at the expense of the AFM peaks. The effect on the $(0,0,1/2)$ peak is to decrease the intensity in the $\pi$ channel to zero without any change to the intensity in the $\sigma$ channel, which remains zero.



\subsection{Magneto-transport}
\begin{figure}[t!]
	\includegraphics[width=0.5\textwidth]{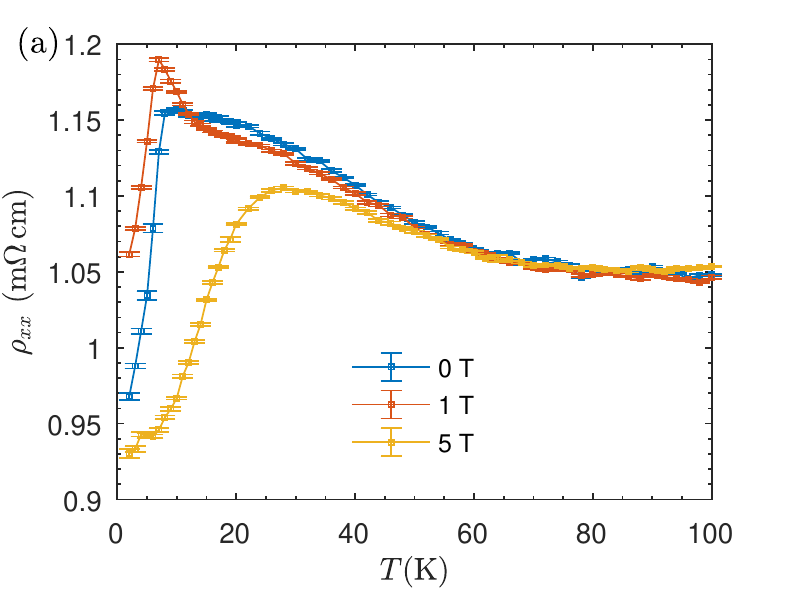}
	\includegraphics[width=0.5\textwidth]{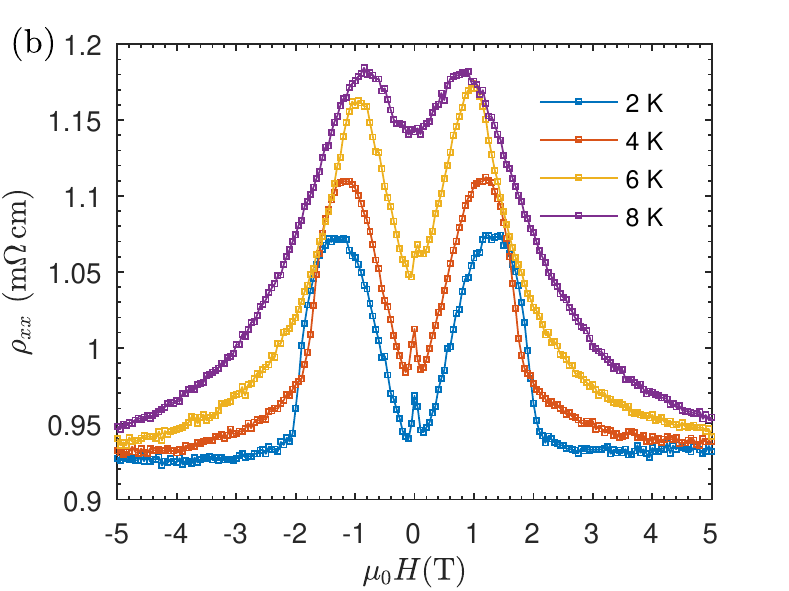}
	\caption{\label{fig:BK2k_EuCd2Sb2}  (a) Temperature dependence of the in-plane resistivity $\rho_{xx}$ at three different field strengths ($H \perp c$). (b) Isothermal $\rho_{xx}(H)$ at several temperatures.}
\end{figure}
Figure~\ref{fig:BK2k_EuCd2Sb2}(a) plots the in-plane resistivity $\rho_{xx}$ as a function of temperature for three different fields. In zero field there is a sharp drop in $\rho_{xx}$ below $T_\textrm{N}$, most likely caused by a suppression of spin scattering due to AFM order.  At $\mu_0H = 1$\,T, $\rho_{xx}$ is slightly peaked around $T_\textrm{N}$, and the reduction below $T_\textrm{N}$ is less than in zero field. Finally, at $\mu_0H = 5$\,T the regime of reduced $\rho_{xx}$ extends up to about 25\,K.


To help understand the magnetoresistance behavior we also measured $\rho_{xx}$ as a function of in-plane field at several temperatures. The data are plotted in Fig.~\ref{fig:BK2k_EuCd2Sb2}(b). At $T = 2$\,K, three distinct features can be identified in  $\rho_{xx}(H)$: (i) In the low field regime ($|\mu_0H| \lesssim$ 0.2 T), associated with the rotation of the spins away from the applied field, there is a small drop in $\rho_{xx}$ which could be due to a reduction in spin scattering as the multiple 120$^\circ$ spin domains form into a single domain. (ii) In the intermediate regime (0.2 T $\lesssim |\mu_0H| \lesssim$ 2 T) there is a peak in $\rho_{xx}$ which ends at the saturation field. The fractional change in $\rho_{xx}(H)$ reaches about 15\% at the maximum. This increase in charge carrier scattering is associated with the canted spin structure, as discussed above. (iii) In the high field regime ($|\mu_0H|> 2$\,T), the spins are fully polarized and the anomalous resistive phase is fully suppressed. At higher temperatures, the initial low-field drop in $\rho_{xx}$ becomes smaller and vanishes for $T > T_\textrm{N}$, and the region of negative magnetoresistance extends to higher fields consistent with the increase in the saturation field with temperature.
\subsection{DFT Calculations}\label{subsec:DFT}

\begin{figure}[b]
	\includegraphics[width=0.5\textwidth]{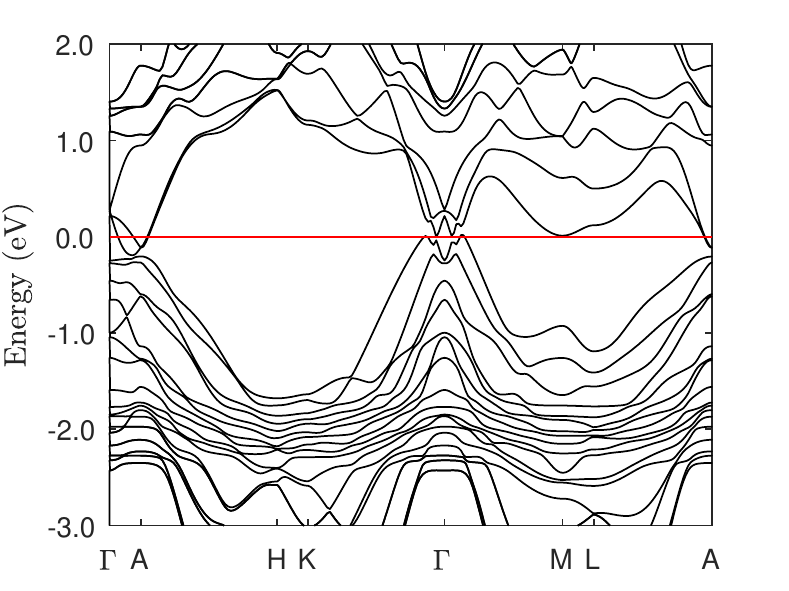}
	\caption{\label{fig:K108a_EuCd2Sb2_bandstructure_Xmgrace_v2} Band structure of EuCd$_2$Sb$_2$ along the high symmetry lines of the hexagonal Brillouin zone~\cite{Setyawan2010}.}
\end{figure}
In this section we discuss the results of \textit{ab initio} electronic structure calculations performed to understand the nature of the electron conduction in the AFM phase of EuCd$_2$Sb$_2$. The calculated electron band structure along high symmetry lines in the Brillouin zone, plotted in  Fig.~\ref{fig:K108a_EuCd2Sb2_bandstructure_Xmgrace_v2}, reveals a low density of electronic states near the Fermi energy, $E_\textrm{F}$. This is consistent with the semimetallic nature of the compound as suggested by transport measurements~\cite{Goryunov2012, Zhang2010}, but contradicts a study of the thermoelectric properties of EuCd$_2$Sb$_2$ by Zhang \textit{et al.}~\cite{Zhang2010}. In that study, a large density of states near $E_\textrm{F}$ was predicted and was attributed to the flat $4f$ electron bands from the europium species residing at the Fermi energy. In fact, the position of these $4f$ bands depends strongly on the value of the Hubbard $U$ parameter. A choice of $U=0$\,eV will cause the $4f$ bands to lie at $E_\textrm{F}$, as found in Ref.~\onlinecite{Zhang2010}. In our calculations we chose a value of $U=3.1$ eV. This choice was guided by a recent ARPES measurement performed on a similar material EuCd$_2$As$_2$ which places the $4f$ electrons $\sim$ 2\,eV below the Fermi level~\cite{Schroeter2016}. On this evidence it is unlikely that the magnetic $4f$ bands contribute significantly to electrical conduction.

\begin{figure}[t]
	\includegraphics[width=0.5\textwidth]{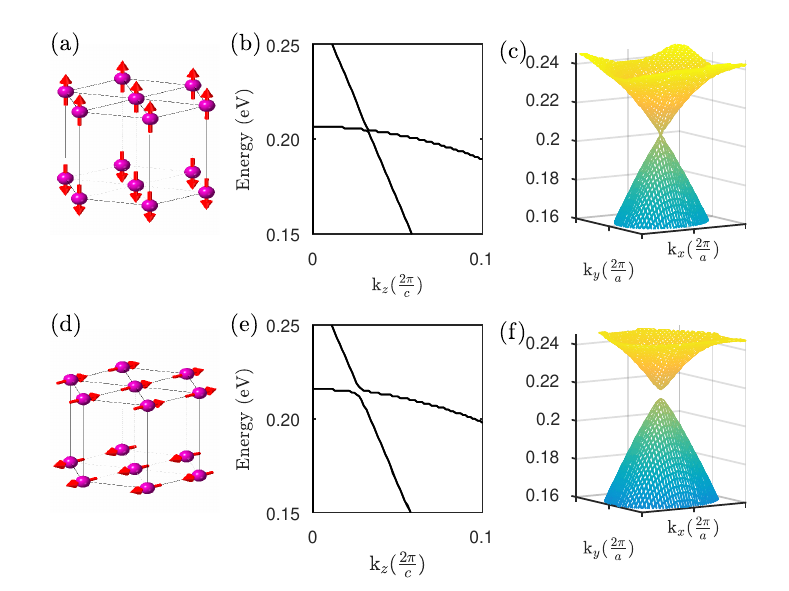}
	\caption{\label{fig:AG11_Dirac_point}  Magnetic structure, detail of bands along the $\Gamma-A$ high symmetry line, and electronic dispersion in the $k_x-k_y$ plane close to the Dirac point. (a)--(c) $\Gamma_3$ magnetic structure. The $\Gamma_3$ irrep of the Eu spin structure would preserve the C$_{3}$ symmetry along $\Gamma-A$ and afford protection to the Dirac point at $k_z=0.0324$. (d)--(f)  $\Gamma_5$ magnetic structure as found here for EuCd$_2$Sb$_2$. The $\Gamma_5$ irrep, on the other hand, breaks this C$_{3}$ symmetry and opens up a gap along $\Gamma-A$ and in the $k_x-k_y$ plane. }
\end{figure}


The Fermi surface comprises one electron pocket and two hole pockets, both with mixed Cd $5s$ and Sb $5p$ orbital character. As mentioned above, the Eu $4f$ states which are responsible for producing the local magnetic moments reside $\sim$ 2 eV below $E_\textrm{F}$ [Fig.~\ref{fig:K108a_EuCd2Sb2_bandstructure_Xmgrace_v2}] (see Supplementary Materials [\onlinecite{Soh2018}]).  This means that the charge carriers in the conducting Cd--Sb double corrugated layer are weakly correlated, and DFT is well suited to calculate the band dispersion in the vicinity of $E_\textrm{F}$.

EuCd$_2$Sb$_2$ has several ingredients  which could lead to non-trivial band topology. First, the heavy masses of Cd and Sb could lead to band inversion due to the large SOC. Second, the three-fold (C$_{3}$) symmetry along the $\Gamma$--$A$ line in reciprocal space [wave vectors $(0,0,k_z)$] might protect some accidental band crossings~\cite{Yang2015}.  Third, in the AFM ordered phase, both non-symmorphic time-reversal symmetry and inversion symmetry of the crystal are preserved,\cite{Hua2018} resulting in a two-fold degeneracy in the electronic bands.

We have identified a gapped band crossing with $\Delta E \sim$ 0.01 eV at $k_z = 0.0266$\,r.l.u. along the $\Gamma-A$ high symmetry line, 0.215\,eV above $E_\textrm{F}$ [Figs.~\ref{fig:AG11_Dirac_point}(e) and (f)]. The energy gap arises because for $T<T_\textrm{N}$ the C$_{3}$ symmetry along the $\Gamma-A$ line is broken by the in-plane orientation of the spins in the $\Gamma_5$ AFM structure. If, instead, the spins were to point along the $c$-axis, as would occur in the $\Gamma_3$ magnetic structure, the linear band crossing would be protected, see Figs.~\ref{fig:AG11_Dirac_point}(b) and (c). In terms of band structure, the bands which cross derive from the $^2$P$_{3/2}$ orbitals of the Sb $5p$ bands with $J_z =1/2$ and $3/2$ (see Supplementary Materials [\onlinecite{Soh2018}]). In a $c$-axis field $J_z$ remains a good quantum number and so the states which cross belong to different irreducible representations of the C$_{3}$ point group. In this case, doubly-degenerate valence and conduction Sb $5p$ bands meet at a four-fold degenerate point in \textbf{k}-space  from which the bands disperse linearly, i.e.~a Dirac point, as depicted in  Figs.~\ref{fig:AG11_Dirac_point}(b) and~\ref{fig:AG11_Dirac_point}(c). In EuCd$_2$Sb$_2$, on the other hand, we have in-plane spin alignment, so the $J_z =1/2$ and $3/2$ bands hybridize, forming a gap at the Dirac point. 

Given the evidence found here for a coupling between Eu spins and electronic conduction it is possible that the magnetically-induced energy gap at the Dirac point could influence the transport properties of  EuCd$_2$Sb$_2$. However, there is no evidence that the gapped band crossing in EuCd$_2$Sb$_2$ is exactly at $E_\textrm{F}$, and there are several other bands in the vicinity of $E_\textrm{F}$ which could contribute to the electronic transport\,(see Supplementary Materials [\onlinecite{Soh2018}]). Nonetheless, our analysis shows that the symmetry in the spin structure controls features in the band structure of EuCd$_2$Sb$_2$, and emphasizes the importance of elucidating the orientation of the Eu spins. It also opens up the possibility of realizing a magnetic Dirac material in the 122-pnictides via spin structure which preserves the C$_{3}$ symmetry. A possible strategy is to find another member of the family  which has spins pointing along the $c$-axis. Indeed a similar situation is found in rare-earth half heuslers where the nature of the band topology is strongly linked to the spin-orientation: Eu$_{0.5}$Ba$_{0.5}$AgBi is predicted to host Weyl fermions protected by the C$_3$ symmetry along the $\Gamma-A$ line where the Eu spins point along the $c$-axis; these Weyl points annihilate in the case of EuAgBi where the spins lie in the basal plane~\cite{Du2015}.

Finally, it is also interesting to compare the physical properties of  EuCd$_2$Sb$_2$ with those of  EuCd$_2$As$_2$~\cite{Wang2016,Hua2018,Rahn2018} and EuZn$_2$Sb$_2$~\cite{weber2006low,C7TA08869H,Zhang2008e}, which are isostructural and also exhibit Eu antiferromagnetic order with similar magnetic ordering temperatures: $9.5$\,K in EuCd$_2$As$_2$ and  $13.3$\,K in EuZn$_2$Sb$_2$. Of these,  EuCd$_2$As$_2$ is a semimetal and EuZn$_2$Sb$_2$ is a semiconductor with a band gap of around 0.5\,eV. A comparison of the band structures reveals a progression in the extent of inversion in the conduction and valence bands. The band inversion is greatest in EuCd$_2$Sb$_2$, with several bands crossing at $E_{\rm F}$ as shown here, whereas in EuZn$_2$Sb$_2$ there is no band crossing. In EuCd$_2$As$_2$ the bands touch at a Dirac point in \textbf{k}-space, which becomes gapped for $T<T_{\rm N}$ for the same reason as described here for EuCd$_2$Sb$_2$. These band crossing features can be understood from the relative sizes of the spin--orbit coupling in the double-corrugated conducting layers, which increases in the order EuZn$_2$Sb$_2$ to EuCd$_2$As$_2$ to EuCd$_2$Sb$_2$. This suggests that a desired level of band crossing can be achieved in the  europium-based 122 pnictide by control of the chemical composition.

\section{Conclusion}
We have determined that the magnetic propagation vector in EuCd$_2$Sb$_2$ is $(0,0,1/2)$, and shown unambiguously that the moments lie in the $(001)$ plane. We have also established how the magnetic structure is changed by an in-plane magnetic field, and find that features observed in the magnetoresistance correlate closely with field-induced changes in the magnetic structure. Our results show that a coupling exists between localized Eu spins and electron transport in EuCd$_2$Sb$_2$. Based on DFT calculations we predict that for $T<T_\textrm{N}$ there exists a gapped Dirac point close to the Fermi level, and although it remains to be seen to what extent this feature influences the charge transport in EuCd$_2$Sb$_2$, our findings suggest that non-trivial band topologies could be induced by magnetic order in the wider family of Eu-based hexagonal 122 pnictides.



\begin{acknowledgments}
The authors  wish  to  thank  M. C. Rahn, C. Topping, H. Jacobsen, N. R. Davies, F. de Juan and P. G. Radaelli for  helpful discussions. We are also grateful to D. C. Johnston (Ames Laboratory) for valuable comments. We thank the University of Oxford's Advanced Research Computing facility (ARCUS-B cluster) for computing resources. This work was supported by the U.K. Engineering and Physical Sciences Research Council, Grant Nos. EP/N034872/1 and EP/M020517/1. J.-R. Soh acknowledges support from the Singapore National Science Scholarship, Agency for Science Technology and Research. 
\end{acknowledgments}

\nocite{*}
\bibliography{library}

\end{document}